\documentclass[aps,prl,showpacs,groupedaddress,twocolumn,preprintnumbers,amsmath,amssymb]{revtex4}

\usepackage{graphicx}% Include figure files
\usepackage{dcolumn}% Align table columns on decimal point
\usepackage{bm}% bold math
\usepackage{braket}

\begin{document}

\title{Spin-Echo-Based Magnetometry with Spinor Bose-Einstein Condensates}% Force line breaks with \\

\author{Yujiro Eto}
\email{eto@qo.phys.gakushuin.ac.jp}
\author{Hayato Ikeda}
\author{Hirosuke Suzuki}
\author{Sho Hasegawa}
\author{Yasushi Tomiyama}
\author{Sawako Sekine}
\author{Mark Sadgrove}
\altaffiliation[Present affiliation: ]{Center for Photonic Innovations, University of Electro-Communications}
\author{Takuya Hirano}
\affiliation{Department of Physics, Gakushuin University, Tokyo 171-8588, Japan}

\date{\today}% It is always \today, today,
             
\begin{abstract}
We demonstrate detection of a weak alternate-current magnetic field by application of the spin-echo technique to $F = 2$ Bose-Einstein condensates.
A magnetic field sensitivity of 12 pT/$\sqrt{\mathrm{Hz}}$ is attained with the atom number of $5 \times 10^{3}$ at spatial resolution of 99 $\mu$m$^{2}$.
Our observations indicate magnetic field fluctuations synchronous with the power supply line frequency.
We show that this noise is greatly suppressed by application of a reverse phase magnetic field.
Our technique is useful in order to create a stable magnetic field environment, which is an important requirement for atomic experiments which require a weak bias magnetic field.
\end{abstract}

\maketitle
Characterization of the inhomogeneity of the ambient magnetic field is an important task in many experiments which utilize atomic systems with spin degrees of freedom.  
Such experiments include fundamental physics such as tests of  symmetries  \cite{Hunter91} and identification of the ground state in spinor condensate systems \cite{Kurn12, Chang04} as well as important applications such as optical lattice clocks \cite{Takamoto05} and long-lived coherence \cite{Langer05}. 
In all of these cases, the experimental accuracy is directly related to the inhomogeneity of the magnetic field.
Recently, high sensitivity magnetometers have been made using both superconducting quantum interference devices and atomic systems.
Although the former are known to provide ultra-sensitive magnetic field sensing, the latter do not require cryogenic cooling and are particularly useful when measuring the ambient magnetic field inside a vacuum chamber.
The most sensitive atomic magnetometer so far realized used a spin-exchange relaxation-free technique to achieve a magnetic field sensitivity of 0.5 fT/$\sqrt{\rm Hz}$ for a measurement volume of 0.3 cm$^3$ \cite{Kominis03}.
However, in such atomic vapour magnetometers the spatial resolution is limited by the diffusion of atoms. 
This problem can be largely overcome by using Bose-Einstein condensates (BECs) in an optical trap. 
For example, it was demonstrated that a sensitivity of 8 pT/$\sqrt{\mathrm{Hz}}$ at spatial resolution of 120 $\mu$m$^2$ could be achieved using an $F = 1$ spinor BEC in Ref. \cite{Vengalattore07}.

All of the atomic systems mentioned above detect constant magnetic fields.
However, specialized magnetometers also exist for detecting alternating current (AC) magnetic fields. 
Such devices have high sensitivity for an AC field within an certain frequency band and are useful for the characterization of temporal fluctuations of the magnetic field.
Such magnetometers can be constructed using the spin-echo technique, as originally demonstrated in experiments on single nitrogen vacancy centers \cite{Taylor08,Maze08,Barasubramanian09}.
In this Letter, we perform a similar type of spin-echo AC magnetometry using $^{87}$Rb $F = 2$ BECs. 
We attain a magnetic field sensitivity of 12 pT/$\sqrt{\mathrm{Hz}}$ with the spatial resolution of about 99 $\mu$m$^{2}$.
To the best of our knowledge this is the first realization of spin-echo based magnetometer using an atomic gas.
In addition, we observe magnetic field noise synchronous with the power supply line at frequencies of 50 and 100 Hz using our AC magnetometer,
and reconstruct the amplitude and phase of the AC magnetic field noise. 
By applying a magnetic field with opposite phase, we can suppress this noise to 1 nT order. We anticipate that the clean magnetic field environment created by this technique will be useful in applications requiring low ambient magnetic fields such as the search for magnetic ground states in systems with spin $> 1/2$ \cite{Kurn12, Chang04} since a temporally fluctuating magnetic fields can cause undesired spin rotations, seriously affecting such experiments. 

%%%% Figure 1%%%%
\begin{figure}[b]
\includegraphics[width=8cm]{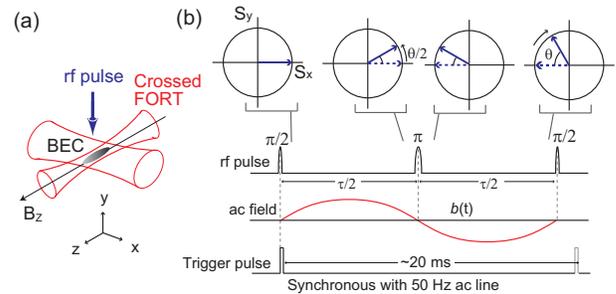}
\caption{(color online) (a) Schematic illustration of the experimental setup.
(b) The upper panel shows the evolution of spin direction in $S_{x}$-$S_{y}$ plane.
The lower panel shows the RF Hahn-echo pulse sequence used for sensing a weak AC magnetic field.
}
\label{fig1}
\end{figure}
%%%%%%%%%%%%%

Figure \ref{fig1}(a) shows the outline of the experimental setup.
A BEC of $^{87}$Rb is created using radio frequency (RF) evaporative cooling in a magnetic trap \cite{Kuwamoto04,Tojo10}.
The BEC is then loaded into a crossed far-off-resonant optical trap (FORT) with axial and radial frequencies of 30 Hz and 100 Hz, respectively.
After 300 ms hold time in the crossed FORT, typically 3$\times$10$^5$ atoms remain in the $|F=2, m_F = -2\rangle$ state.
A bias magnetic field along the $z$ axis ($B_z$) shown in Fig. 1(a) is applied to define the quantization axis.
To generate the stable $B_z$ field, a laser diode source with a low ripple noise of less than 2 $\mu$A (Newport 505) is used as the current source for our $z$-axis Helmholtz coils and the whole experimental setup is installed inside a magnetic shield room whose walls consist of permalloy plates.
The magnetic field along the $x$- and $y$-direction is carefully compensated by using two Helmholtz coils with the similar laser current sources. 
The strength of the $B_z$ field is calibrated from the Larmor frequency as measured using a Ramsey interferometric method \cite{Mark13} and is found to be  $B_{z} =$ $9.26$ $\mu$T for a Helmholtz coil current of 120 mA.

%%%% Figure 2%%%%
\begin{figure}[t]
\includegraphics[width=7cm]{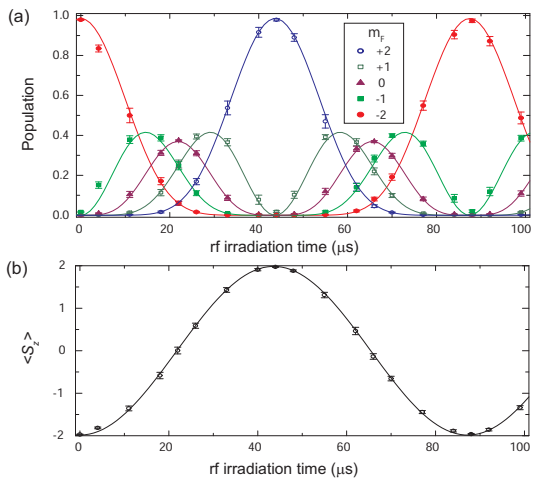}
\caption{(color online) Rabi-type oscillation of our spin 2 BEC.
(a) and (b) show the population of each $m_{F}$ component  ($N_{m_{F}}/\Sigma_{m_{F}=-2}^{+2} N_{m_{F}}$) and $\langle S_{z} \rangle$ as a function of the RF irradiation time, respectively.
The RF pulse envelope is shaped by a Gaussian function and the irradiation time corresponds to twice the standard deviation.
Each plot represents the average over the ten measurements.
The solid curves in (a) and (b) are fitting curves obtained using the spin-2 rotation matrix \cite{Mark13} and the cosine function, respectively.
}
\label{fig2}
\end{figure}
%%%%%%%%%%%%%

The time sequence used for sensing the weak magnetic field is shown in the bottom panel of Fig. \ref{fig1}(b). 
In the detection of the weak AC magnetic field using the spin echo technique  \cite{Taylor08},
the RF Hahn-echo pulse sequence ($\pi/2$-$\pi$-$\pi/2$) is applied to the initial $| 2, -2 \rangle$ state in the crossed FORT.
The first $\pi/2$ pulse rotates the spin vector from the $S_z$ direction to the $S_x$ direction and induces Larmor precession in the $S_x-S_y$ plane as shown in Fig. 1(b). 
We synchronize the RF source with the AC power line so that the timing of the first $\pi/2$ pulse is synchronous with the 50 Hz supply.
The upper panel in Fig. 1(b) shows the evolution of the spin vector in the $S_{x}$-$S_{y}$ plane between two $\pi$/2 pulses,
where the frame is rotating at Larmor frequency of $f_{0} = g_{F} \mu_{B} B_{z}/h$ with $g_{F}$ and $\mu_{B}$ being g-factor and the Bohr magneton.
When we apply a time-varying magnetic filed $b(t)$ along the $z$-direction, the spin direction in the $S_{x}$-$S_{y}$ plane is changed by $\theta = \frac{g_{F} \mu_{B}}{\hbar}[\int^{\tau/2}_{0} b(t) dt-\int^{\tau}_{\tau/2} b(t) dt]$ relative to the $-S_x$ direction due to the presence of the AC magnetic field $b(t)$ (shown by solid arrows in the top panel of Fig. 1(b)).
Note that the maximum $\theta$ variation for a single frequency field, $b(t) = b_{\mathrm{AC}} \sin{(2\pi t/\tau_{\mathrm{AC}} )}$, is reached when $\tau$, the total precession time between two $\pi/2$ pulses, is equal to $\tau_{\mathrm{AC}}$.
The angle of $\theta$ yielded by $b(t)$ is converted to $S_{z}$, by the application of a second $\pi/2$ pulse,
and the relationship between the expectation value of $S_{z}$, $\langle S_{z} \rangle$, and $b(t)$ can be expressed by 
\begin{equation}
\langle S_{z}\rangle = -2 \cos{\{ \frac{g_{F} \mu_{B}}{\hbar}[\int^{\tau/2}_{0} b(t) dt-\int^{\tau}_{\tau/2} b(t) dt]\}}.
\label{eq1}
\end{equation}
In the case that $b_{\mathrm{AC}}=0$ (shown by dotted arrows in the top panel of Fig. 1(b)), the spin direction returns to the  $| 2, -2 \rangle$ state ($\langle S_{z}\rangle=-2$).
One can thus detect the weak AC magnetic field by measuring the $\langle S_{z} \rangle$.
In addition, the effect of undesirable inhomogeneities such as magnetic field gradients and slowly fluctuating magnetic fields is reduced due to spin-echo \cite{Eto13}.

The techniques of the Stern-Gerlach separation and time-of-flight absorption imaging are used to obtain $\langle S_{z} \rangle$.
The atomic density distributions of each $m_{F}$ component are measured by shining the imaging beam from the $x$-direction after a time of flight of 15 ms.
The atom number in each $m_{F}$ component, $N_{m_{F}}$, is calculated over a small region in the center of the BEC (66 $\mu$m and 47 $\mu$m in the $y$- and $z$-direction) in order to extract the peak atomic number.  
Using $N_{m_{F}}$, we calculate $\langle S_{z} \rangle$ from the following equation: $\langle S_{z} \rangle = \Sigma_{m_{F}=-2}^{+2} m_{F} N_{m_{F}}/\Sigma_{m_{F}=-2}^{+2} N_{m_{F}}$.

In the first experiment, we observed Rabi-type oscillations of the spin-2 BEC in order to determine the RF pulse durations for $\pi/2$ and $\pi$ pulses.
Instead of the Hahn echo sequence, a single RF pulse with various irradiation time was applied to $| 2, -2 \rangle$ state.
Figure \ref{fig2}(a) and 2(b) show the population of each $m_{F}$ component ($N_{m_{F}}/\Sigma_{m_{F}=-2}^{+2} N_{m_{F}}$) and  $\langle S_{z} \rangle$ as a function of the RF irradiation time, respectively.
The clear spin 2 rotation was observed, and we found that the irradiation of 21.8 and 43.6 $\mu$s correspond to the $\pi/2$ and $\pi$ pulse.

\begin{figure}[t]
\includegraphics[width=7cm]{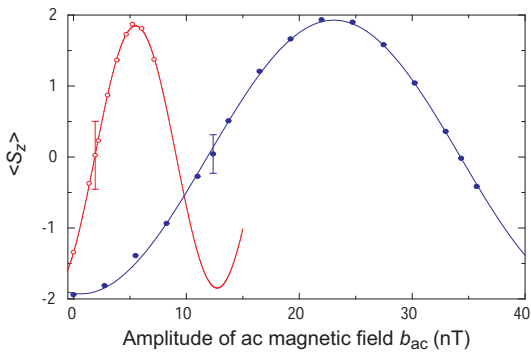}
\caption{(color online) Operation of the AC magnetometer using the Hahn echo sequence.
The $\langle S_{z} \rangle$ values were measured as a function of $b_{\mathrm{\mathrm{AC}}}$, where a single frequency AC magnetic field $b(t)$ is applied.
The filled and open circles indicate the measured data at $\tau$ = 5 and 15 ms, respectively.
The plot at $\langle S_{z} \rangle \sim 0$ in Fig. 2 is the average over 60 measurements, and the other plots are the average of 30 measurements.
}
\label{fig3}
\end{figure}

In order to confirm that our system operates as an AC magnetometer,
$\langle S_{z} \rangle$ was measured in the presence of a purposely introduced AC magnetic field of amplitude $b_{\mathrm{AC}}$.
Figure \ref{fig3} shows the measured value of $\langle S_{z} \rangle$ versus $b_{\mathrm{AC}}$ at $\tau = 5$ ms (filled circles) and 15 ms (empty circles),
where $b(t) = b_{\mathrm{AC}} \sin{(2\pi t/\tau)}$ and the Hahn echo sequence is also applied.
The solid curves in Fig. \ref{fig3} are cosine functions fitted to the data where the amplitude, period and initial phase were the fitting parameters.
The cosinusoidal variation shown in Fig. \ref{fig3} means that the phase variation in the $S_{x}$-$S_{y}$ plane accumulated by the AC magnetic field was successfully observed.
The period at $\tau = $15 ms is three times shorter than that at 5 ms due to the longer phase accumulation time.

%%%% Figure 3%%%%
%%%%%%%%%%%%%

%%%% Figure 4%%%%
\begin{figure}[b]
\includegraphics[width=7cm]{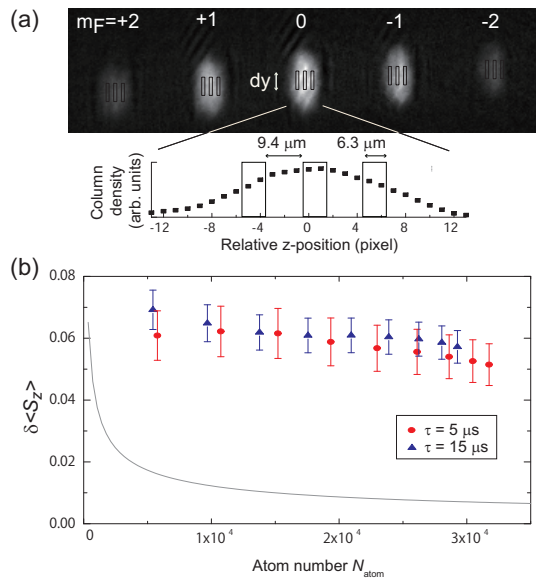}
\caption{(color online) (a) The upper panel shows a typical optical density distribution measured at $\langle S_{z} \rangle \sim 0$.
The lower panel shows the column density, where the optical density along $y$-direction is integrated. 
The optical density distribution of each $m_{F}$ component is divided into three regions.
Taking into account the finite resolution of our imaging system ($\sim$ 7.5 $\mu$m) \cite{Eto13}, each region is separated by 9.4 $\mu$m.
(b) $\delta \langle S_{z} \rangle$ as a function of the total atom number averaged over three regions, $N_{\mathrm{atom}}$, where the $N_{\mathrm{atom}}$ is changed by increasing the region along the $y$-direction, $dy$.
The solid curve indicates the variance induced by the atom shot noise. 
}
\label{fig4}
\end{figure}
%%%%%%%%%%%%%

The magnetometer sensitivity for a single measurement is given by  $\delta b_{\mathrm{min}} = \delta \langle S_{z} \rangle/(d \langle S_{z} \rangle/d b_{\mathrm{AC}})$, 
where  $\delta \langle S_{z} \rangle$ is the standard deviation of $\langle S_{z} \rangle$ in a single measurement and $d \langle S_{z} \rangle/d b_{\mathrm{AC}}$ is the slope with $b_{\mathrm{AC}}$.
When we calculate the $\delta \langle S_{z} \rangle$ from the measured $\langle S_{z} \rangle$ values at $\langle S_{z} \rangle \sim 0$,
$\delta \langle S_{z} \rangle$ is found to be 0.28 and 0.48 at $\tau = 5$ and $15$ ms, respectively.
From these values, we find the sensitivity to be $\delta b_{\mathrm{min}} = 0.97$ and $0.66$ nT.
Note that the spin echo technique can only remove the effect of slowly fluctuating magnetic fields whose period is longer than $\tau$.
However, our experiment is subject to the influence of faster fluctuations that vary for each measurement. 
For example, the magnetic field noise caused by the ripple of the Helmholtz coil current is of order 0.1 nT.

In order to evaluate the intrinsic sensitivity, which is unaffected by the temporal fluctuation of the magnetic field,
we divide the optical density distribution of each $m_{F}$ component for a single measurement into the three regions as shown in Fig. \ref{fig4}(a).
We infer the value of $\delta \langle S_{z} \rangle$ from three expectation values of $S_{z}$, $\langle S_{z} \rangle_{i} = \Sigma_{m_{F}=-2}^{+2} m_{F} N_{i, m_{F}}/\Sigma_{m_{F}=-2}^{+2} N_{i, m_{F}}$,
where the subscript $i=1-3$ indicates which of the three regions is being considered.
In our previous work \cite{Eto13}, 
we theoretically and experimentally confirmed that the shape of the atomic distribution of each $m_{F}$ component in the optical trap is almost unchanged after a time-of-flight of 15 ms,
although the distributions become more spread out.
Each value of $\langle S_{z} \rangle_{i}$ 
thus reflects the value of $\langle S_z \rangle$ in a different \emph{spatial} region of the trapped BEC.
A field sensitivity of $94 \pm 9$ pT for $\tau=15$ ms is attained with corresponding  $\delta \langle S_{z} \rangle$ of $0.069\pm0.006$,
when we select the size of each region as  $dy \times dz = 15.7$ $\mu$m $\times$ 6.3  $\mu$m = 99 $\mu$m$^{2}$, where
$dy$ and $dz$ represent the length along the $y$- and $z$-direction of each region. 
%The spatial resolution of our magnetometer thus has $99$ $\mu$m$^{2}$ at least.
The field sensitivity for $N$ times measurement per 1 second is to be $\delta  b_{\mathrm{min}}^{N} = \delta  b_{\mathrm{min}}/\sqrt{1/(15 \times10^{-3})} = 12\pm1$ pT/$\sqrt{\mathrm{Hz}}$.  

Figure \ref{fig4}(b) shows $\delta \langle S_{z} \rangle$ versus atom number averaged over three regions, $N_{\mathrm{atom}}=\Sigma_{i=1}^{3}\Sigma_{m_{F}=-2}^{+2} N_{i, m_{F}}/3$, 
where $N_{\mathrm{atom}}$ is changed by increasing $dy$.
The dotted curve represents the atom shot noise limited $\delta \langle S_{z} \rangle$.
The deviation from the the atom shot noise limited values has multiple origins which induce spatial distortion of the atomic distributions and the optical image: 
Interference fringes caused by unclean regions on the imaging optics is one well known origin of distortion in the image.
Additionally, the effect of the magnetic field gradient ($\sim1.5$ $\mu$T/cm), which cannot be completely removed by spin-echo  \cite{Yasunaga08}, and spontaneous pattern formation \cite{Kronjager10} produce effective distortion of the atomic distribution.
Another possible cause of the deviation is the effect of thermal atoms, whose Gaussian tails reduce the accuracy of discrimination between the $m_F$ components in the Stern-Gerlach separation. 
Thus the sensitivity of our AC magnetometer will be improved by reduction of the magnetic field gradient and optimization of imaging system. 

\begin{figure}[t]
\includegraphics[width=7cm]{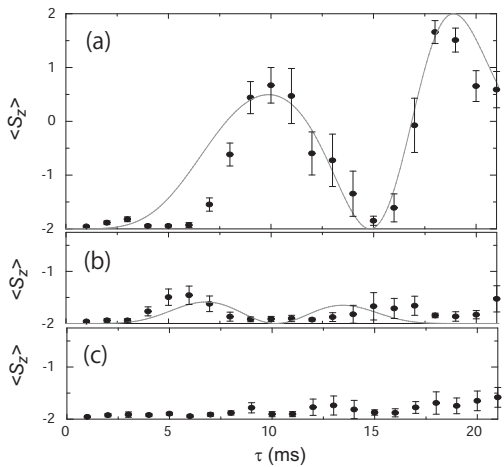}
\caption{$\tau$ dependence of $\langle S_{z} \rangle$ values after the Hahn echo sequence.
Each point represents the average over ten measurements with the error bars giving the standard deviation over those measurements.
(a) $\langle S_{z} \rangle$ values measured without the artificial application of the AC magnetic field.
(b) $\langle S_{z} \rangle$ values measured with application of an inverse phase magnetic field at 50 Hz, $b(t) = b_{20} \sin{[2 \pi t/(20 \times 10^{-3})+\theta_{20}]}$,
where $b_{20} = 9.5$ nT and $\theta_{20} = -0.76+\pi$.
(c) $\langle S_{z} \rangle$ values measured with application of an inverse phase magnetic field at 50 Hz and 100 Hz, $b(t) = b_{20} \sin{[2 \pi t/(20 \times 10^{-3})+\theta_{20}]}+ b_{10} \sin{[2 \pi t/(10 \times 10^{-3})+\theta_{10}]}$,
where $b_{20} = 3.4$ nT and $\theta_{20} = 1.50+\pi$.
}
\label{fig5}
\end{figure}

As a practical application of our magnetometer, we performed detection of the stray AC magnetic field present in our apparatus.
Figure \ref{fig5}(a) shows $\langle S_{z} \rangle$ values measured as a function of $\tau$,
without artificial AC magnetic field.
If the stray AC magnetic field in the region occupied by the BEC fluctuates in a random manner with respect to amplitude, 
frequency and phase, then we would expect that the observed values of $\langle S_z \rangle$ would also exhibit a random distribution. 
Instead, we see that $\langle  S_z \rangle$ exhibits oscillatory behavior. 
Such behavior indicates the existence of a stable AC stray magnetic field in our apparatus.
Note that the first $\pi/2$ pulse in the Hahn echo sequence is synchronous with the power supply line of 50 Hz.
It is therefore reasonable to expect that the AC stray magnetic field is mainly induced by the magnetic field arising from the electronic devices surrounding the BEC apparatus which should be synchronous with the 50 Hz supply line.

%%%% Figure 5%%%%
%%%%%%%%%%%%%

In order to quantitatively investigate the effect of  AC stray magnetic field,
we simulate the  $\tau$ dependence of $\langle S_{z} \rangle$ by inserting a 50 Hz magnetic field, $b(t) = b_{20} \sin{[2 \pi t/(20 \times 10^{-3})+\theta_{20}]}$, into the Eq. 1.
The solid curve in Fig.  \ref{fig5}(a) represents a fit of Eq. 1 to the data assuming such a 50 Hz magnetic field,
where $b_{20}$ and $\theta_{20}$ are used as the fitting parameters.  
The fitted curve reproduces the oscillating behavior of the data in Fig. \ref{fig5}(a),
and the parameter values are found to be $b_{20} = 9.5$ nT and $\theta_{20} = -0.76$.

Assuming that the fitted parameter values are accurate, we should be able to remove the effects of the 50 Hz stray magnetic field along $z$-direction by application of a magnetic field with the same frequency and amplitude but inverse phase.
Figure \ref{fig5}(b) shows $\langle S_z \rangle$ measured for the same parameters as in Fig. 5(a) but in the presence of an artificially applied 50 Hz magnetic field with $b_{20}=$ 9.5 nT and $\theta_{20}=-0.76+\pi$.
The variation of  $\langle S_{z} \rangle$ values are clearly suppressed compared with Fig. \ref{fig5}(a),
and this suppression suggests that the 50 Hz magnetic field noise is reduced.

Nonetheless, the small remaining oscillation in Fig. \ref{fig5}(b) indicates that a synchronous magnetic field still remains along the $z$-direction. 
We assumed the existence of a 100 Hz magnetic field, $b_{10}(t) = b_{10} \sin{[2 \pi t/(10 \times 10^{-3})+\theta_{10}]}$,  and by fitting to the data in Fig. 5(b) we obtained the parameters $b_{10}=3.4$ nT and $\theta_{10} = 1.50$.
Based on the parameter values, 
we further applied an inverse phase magnetic field at 100 Hz in addition to the application of that at 50 Hz [Fig. \ref{fig5}(c)].
As shown in Fig. \ref{fig5}(c), 
$\langle S_z \rangle$ is close to $-2$ for most values of $\tau$, particularly when $\tau\leq 10$ ms.
This result indicates that magnetic field noise which is synchronous with the power supply line is strongly suppressed by the application of  an inverse phase field with frequency components at 50 and 100 Hz. 
In particular, we note that at $\tau=20$ ms, 
the value of $\langle S_z \rangle$ is $-1.7\pm0.2$. 
This value is consistent with an AC magnetic field $b(t)=b_{20}\sin[2\pi t/(20\times10^{-3})]$ with $b_{20} = 1.1^{+0.2}_{-0.4}$ nT, implying that the magnetic field has been suppressed by almost one order of magnitude when compared with the uncompensated case shown in Fig. \ref{fig5}(a).

In conclusion we have reported the demonstration of a spin-echo based magnetometer using $^{87}$Rb $F = 2$ Bose-Einstein condensates,
with which weak AC magnetic field noise is detected.
We attained a field sensitivity of $12$ pT/$\sqrt{\rm Hz}$ at a spatial resolution of 99 $\mu$m$^2$.
In addition we observed magnetic field noise synchronous with the power supply line.
By artificial application of an inverse phase AC magnetic filed, 
the synchronous noise at 50 Hz is suppressed down to 1 nT order. 
The techniques demonstrated here are use for characterizing and controlling the magnetic field environment, with particular applicability to ultracold atom experiments. 
By allowing the detection and creation of a stable, weak bias magnetic field we anticipate that our magnetometer will facilitate the development of fundamental research areas in atomic physics which require a weak magnetic field regime \cite{Hunter91,Kurn12,Chang04,Takamoto05}.

We would like to thank T. Kuwamoto for fruitful discussions.
This work was supported by the Japan Society for the Promotion of Science (JSPS) through its Funding Program for World-Leading Innovation R\&D on Science and Technology (FIRST Program).

\end{document}